\documentclass[aps, amsmath, amssymb,reprint,superscriptaddress]{revtex4-2}

\usepackage[utf8]{inputenc}
\usepackage{graphicx}
\usepackage{dcolumn}
\usepackage{bm}
\usepackage[T1]{fontenc}
\usepackage{mathptmx}
\usepackage{gensymb}
\usepackage{upgreek}
\usepackage{xcolor}
\usepackage{textcomp}
\usepackage{hyperref}
\usepackage{mhchem}
\usepackage{float}
\usepackage[verbose]{placeins} 
\hypersetup{colorlinks=true,linkcolor=red,citecolor=blue,filecolor=magenta,urlcolor=magenta}
\usepackage{setspace}
\usepackage{etoolbox}

\begin{document}


\title{On-Chip Generation of Co-Polarized and Spectrally Separable Photon Pairs}

\author{Xiaojie Wang}
\affiliation{Department of Materials Science and Engineering, National University of Singapore, 117575, Singapore}
\affiliation{Centre for Quantum Technologies, National University of Singapore, 117543, Singapore}

\author{Lin Zhou}%
\affiliation{Department of Materials Science and Engineering, National University of Singapore, 117575, Singapore}
\affiliation{Centre for Quantum Technologies, National University of Singapore, 117543, Singapore}

\author{Yue Li}%
\affiliation{Department of Materials Science and Engineering, National University of Singapore, 117575, Singapore}

\author{Sakthi Sanjeev Mohanraj}%
\affiliation{Department of Materials Science and Engineering, National University of Singapore, 117575, Singapore}

\author{Xiaodong Shi}
\affiliation{A$^\ast$STAR Quantum Innovation Centre (Q.InC), Agency for Science, Technology and Research (A$^\ast$STAR), 138634, Singapore}%
\affiliation{Institute of Materials Research and Engineering (IMRE), Agency for Science, Technology and Research (A$^\ast$STAR), 138634, Singapore}%

\author{Zhuoyang Yu}%
\affiliation{Department of Materials Science and Engineering, National University of Singapore, 117575, Singapore}
\affiliation{Centre for Quantum Technologies, National University of Singapore, 117543, Singapore}


\author{Ran Yang}%
\affiliation{Department of Materials Science and Engineering, National University of Singapore, 117575, Singapore}

\author{Xu Chen}%
\affiliation{Department of Materials Science and Engineering, National University of Singapore, 117575, Singapore}

\author{Guangxing Wu}%
\affiliation{Department of Materials Science and Engineering, National University of Singapore, 117575, Singapore}
\affiliation{Centre for Quantum Technologies, National University of Singapore, 117543, Singapore}

\author{Hao Hao}
\affiliation{Centre for Quantum Technologies, National University of Singapore, 117543, Singapore}

\author{Sihao Wang}
\affiliation{A$^\ast$STAR Quantum Innovation Centre (Q.InC), Agency for Science, Technology and Research (A$^\ast$STAR), 138634, Singapore}%
\affiliation{Institute of Materials Research and Engineering (IMRE), Agency for Science, Technology and Research (A$^\ast$STAR), 138634, Singapore}%

\author{Veerendra Dhyani}
\affiliation{A$^\ast$STAR Quantum Innovation Centre (Q.InC), Agency for Science, Technology and Research (A$^\ast$STAR), 138634, Singapore}%
\affiliation{Institute of Materials Research and Engineering (IMRE), Agency for Science, Technology and Research (A$^\ast$STAR), 138634, Singapore}%

\author{Di Zhu}
\email{dizhu@nus.edu.sg}
\affiliation{Department of Materials Science and Engineering, National University of Singapore, 117575, Singapore}
\affiliation{Centre for Quantum Technologies, National University of Singapore, 117543, Singapore}
\affiliation{A$^\ast$STAR Quantum Innovation Centre (Q.InC), Agency for Science, Technology and Research (A$^\ast$STAR), 138634, Singapore}
\affiliation{Institute of Materials Research and Engineering (IMRE), Agency for Science, Technology and Research (A$^\ast$STAR), 138634, Singapore}

\begin{abstract}
On-chip generation of high-purity single photons is essential for scalable photonic quantum technologies. Spontaneous parametric down-conversion (SPDC) is widely used to generate photon pairs for heralded single-photon sources, but intrinsic spectral correlations of the pairs often limit the purity and interference visibility of the heralded photons. Existing approaches to suppress these correlations rely on narrowband spectral filtering, which introduces loss, or exploiting different polarizations, which complicates on-chip integration. Here, we demonstrate a new strategy for generating spectrally separable photon pairs in thin-film lithium niobate nanophotonic circuits by harnessing higher-order spatial modes, with all interacting fields residing in the same polarization. Spectral separability is achieved by engineering group-velocity matching using higher-order transverse-electric modes, combined with a Gaussian-apodized poling profile to further suppress residual correlations inherent to standard periodic poling. Subsequent on-chip mode conversion with efficiency exceeding 95\% maps the higher-order mode to the fundamental mode and routes the photons into distinct output channels. The resulting heralded photons exhibit spectral purities exceeding 94\% inferred from joint-spectral intensity and 89\% from unheralded $g^{(2)}$ measurement. This approach enables flexible spectral and temporal engineering of on-chip quantum light sources for quantum computing and quantum networking.
\end{abstract}

\maketitle

\section{Introduction}

Integrated photonics offers a viable path towards large-scale quantum information processing~\cite{madsen2022quantum,flamini2018photonic,multiphoton,psiquantum2025}, enabling compact, phase-stable, and manufacturable hardware systems. A key building block for a photonic quantum processor is a stable source of indistinguishable single photons\cite{HOM1987}, whose interference underpins basic quantum logic operations~\cite{2019photonic,kok2007linear,zhong2020quantum}. Spontaneous parametric down-conversion (SPDC) is one of the most widely adopted methods to generate single photons due to its reliability, controllability, and reproducibility. In SPDC processes, a pump photon inside a second-order nonlinear medium is down-converted into a pair of lower-energy photons, referred to as the signal and idler. Detecting one heralds the presence of the other\cite{2008PRLwal,2017PRLheralded}. However, photons produced by SPDC often carry unwanted spectral correlations due to the intrinsic energy conservation requirement. These correlations reduce spectral purities of the heralded photons, limiting the interference visibility in subsequent multi-photon quantum operations~\cite{2001PRAeliminating}. While narrowband optical filtering can reduce the spectral correlation and improve purity, it inevitably introduces loss and lowers the brightness of the source~\cite{2017PRALimits}. It remains an outstanding challenge to realize high-performance, integrated SPDC sources that produce spectrally pure photons without resorting to narrowband filtering. 

Significant efforts have been devoted to addressing this challenge~\cite{on-chip-sources,APR2025nanodomain,2018design}. One way is to use cavities, which achieve high purity and indistinguishability by imposing narrow optical resonances\cite{PRL2020ultrabright,2023pranarrow}, but the photon bandwidth is limited, and the devices are sensitive to resonance fluctuations. The other way is to use dispersion-engineered waveguides, which shape the group-velocity relations among interacting fields to suppress spectral correlations. Two main configurations are commonly used. The first—and most widely adopted—is type-II SPDC, where the signal and idler photons have orthogonal polarizations with different dispersions and can therefore satisfy the group-velocity-matching (GVM) condition\cite{1997spectral,2025typeii,xin2022}. While effective, type-II implementations require polarization management\cite{2019SA}, including rotators, splitters, and combiners, which complicates photonic circuit design and may introduce additional losses. The second approach is backward-wave SPDC, in which the signal and idler propagate in opposite directions\cite{2025counter,2021counter}. This method can generate exceptionally narrowband photons, often one to two orders of magnitude narrower than those from conventional co-propagating SPDC\cite{2018heralding}. However, this method requires submicrometer poling period, making fabrication particularly challenging.

\begin{figure*}
    \centering
    \includegraphics[width=0.9\linewidth]{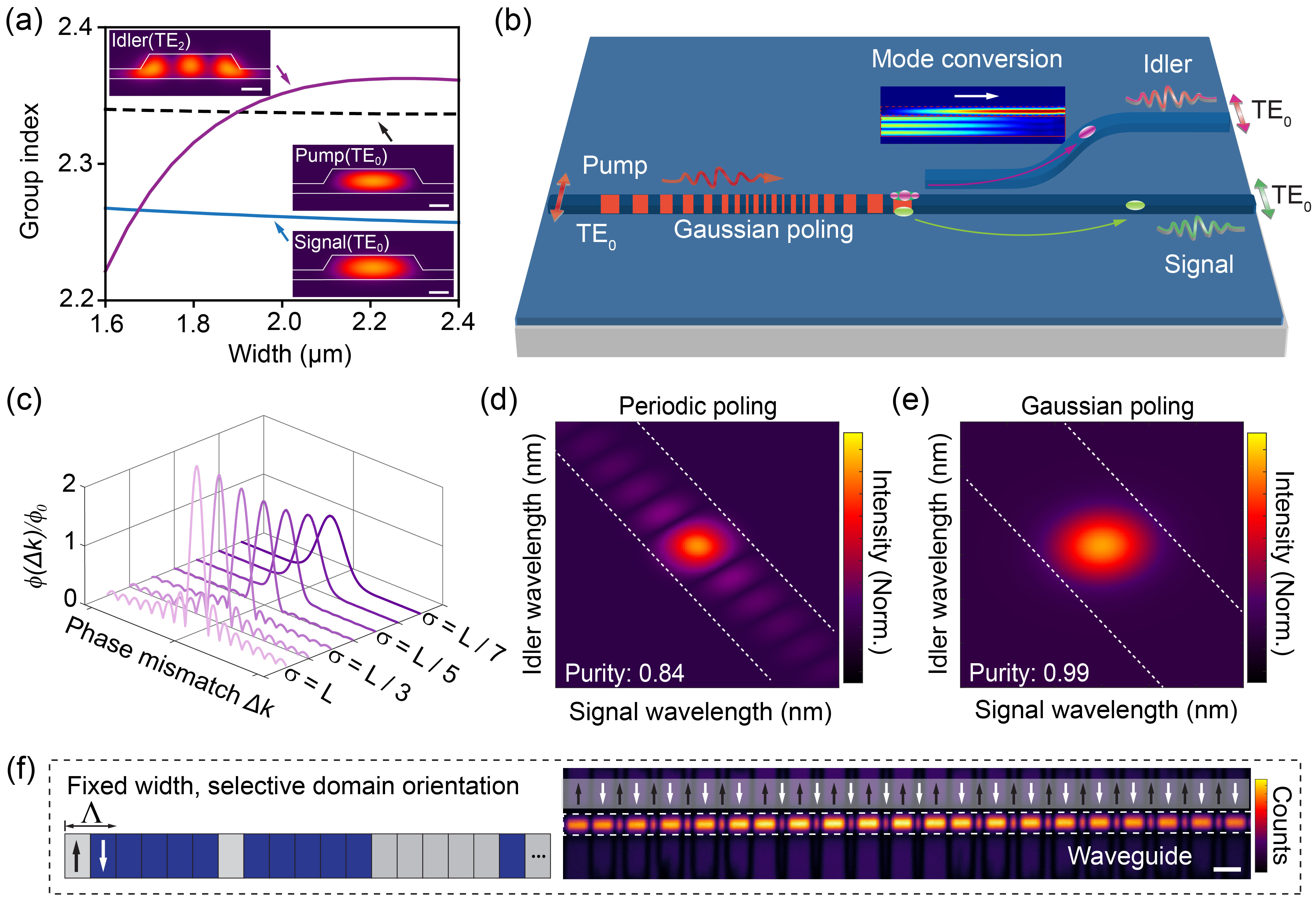}
    \caption{\label{Figure1}
    Design principle and simulation of the co-polarized spectrally separable photon pair (SSPP) source. 
    (a) Simulated group indices and corresponding optical mode profiles of the pump (785 nm, TE$_0$), signal (1520 nm, TE$_0$), and idler (1620 nm, TE$_2$) in the TFLN waveguide, illustrating the modal configuration used in this work. Scale bar: 500 nm.
    (b) Schematic of the integrated device incorporating the Gaussian-apodized poling region and the on-chip mode converter.
    (c) Target PMF showing the balance between spectral purity and effective nonlinear strength (brightness).
    (d) Simulated JSA with periodic poling, showing a sinc-shaped PMF with residual spectral correlations in the sidelobes.
    (e) Simulated JSA with Gaussian-apodized poling, yielding a Gaussian spectrum with suppressed sidelobes.
    (f) Schematic of the Gaussian-apodized domain inversions (left) and SHG microscope image confirming the realized domain inversions (right). Scale bar: $5~\mu\mathrm{m}$.
    }
\end{figure*}

In this work, we demonstrate co-polarized, spectrally separable photon pairs (co-polarized SSPP) generated on thin-film lithium niobate (TFLN), alleviating polarization-handling requirements and simplifying photonic circuit design. The TFLN platform offers flexible domain-engineering capabilities\cite{wangcoptica2018}, strong electro-optic tunability\cite{zhudilight,zhudireview}, and an increasingly mature ecosystem for integrated nonlinear and quantum photonics\cite{2020PRL}. The key idea of our approach is to leverage higher-order TE modes to introduce an additional knob for dispersion engineering, satisfying GVM condition with a single polarization under type-0 phase matching. To interface these modes with standard photonic circuits, an on-chip mode converter maps the higher-order mode back to the fundamental mode (TE$_0$) and routes the signal and idler photons into distinct paths. In parallel, a Gaussian-apodized poling pattern reshapes the nonlinear response and suppresses residual spectral correlations, enabling a measured spectral purity of 94\% without filtering. In contrast to type-II configurations, where phase matching is constrained by the relatively rigid dispersion of cross-polarized modes, higher-order modes offer stronger dispersion tunability. This tunability enables flexible control of photon temporal and spectral modes in integrated quantum sources.

\section{Results}
\subsection{Device design}
The SPDC process mediates the conversion of a high-energy pump photon into a pair of lower-energy signal and idler photons. The biphoton component of the state can be expressed as
\begin{equation}
|\psi \rangle = \int\int d\omega_s d\omega_i f(\omega_s,\omega_i)
\hat{a}_s^\dagger(\omega_s)\hat{a}_i^\dagger(\omega_i)|0\rangle ,
\end{equation}
where $f(\omega_s,\omega_i)$ is the JSA of the generated photon pairs.
To obtain spectrally separable photon pairs, the JSA must be factorable as $f(\omega_s,\omega_i)=f(\omega_s)f(\omega_i)$. It can be expressed as the product of the pump envelope function (PEF), $\alpha(\omega_s+\omega_i)$, and the phase-matching function (PMF), $\phi(\omega_s,\omega_i)$\cite{2006generation},
\begin{equation}
f(\omega_s,\omega_i)=\alpha(\omega_s+\omega_i)\phi(\omega_s,\omega_i),
\end{equation}

and \begin{equation}
\phi(\omega_s,\omega_i)=\int_0^{L} g(z)e^{i\Delta k(\omega_s,\omega_i)z}dz,
\end{equation}
where $g(z)=\chi^{(2)}(z)/\chi_0^{(2)}$ represents the normalized spatial modulation of the nonlinear coefficient, $L$ denotes the interaction length, $\Delta k=k_p-k_s-k_i$ is the phase mismatch. Expanding the wave vectors near ($\omega_{i0}$, $\omega_{s0}$) yields
\begin{equation}
\begin{aligned}
\Delta k = \Delta k_0 
&+ (v_p^{-1}-v_i^{-1})(\omega_i-\omega_{i0})+ (v_p^{-1}-v_s^{-1})(\omega_s-\omega_{s0}),
\end{aligned}
\end{equation}
where $\omega_{s0}$ and $\omega_{i0}$ are the signal and idler central frequencies,
$v_p$, $v_s$, and $v_i$ are the group velocities of the pump, signal, and idler photons, and 
$\Delta k_0$ is the phase mismatch calculated at the central frequencies.
To realize a separable JSA, the PEF and PMF should be properly oriented in the $(\omega_s, \omega_i)$ plane. The PEF, determined by energy conservation, is always antidiagonal, whereas the PMF lies along an axis defined by the group velocities of the pump, signal, and idler fields as
\begin{equation}
\tan\theta = -\frac{v_p^{-1} - v_s^{-1}}{v_p^{-1} - v_i^{-1}}.
\label{eq:GVM}
\end{equation}
where $\tan\theta$ defines its slope in the $(\omega_s, \omega_i)$ plane. By engineering the group velocities of the interacting modes\cite{mosley2008conditional, 2006generation}, the PMF can be rotated to become orthogonal to the PEF, fulfilling the GVM condition and yielding a nearly factorable JSA with strongly suppressed spectral correlations.

Building on this theoretical framework, we employ higher-order spatial modes to engineer the modal dispersion in a type-0 phase-matching configuration. In our design, both the pump and the signal propagate in the $\mathrm{TE_0}$ mode, while the idler is generated in the higher-order $\mathrm{TE_2}$ mode. As shown in Fig.~1a, when the width of the waveguide is increased to above ~2 µm with an etch depth of 360 nm, the dispersion curves of the $\mathrm{TE_0}$ and $\mathrm{TE_2}$ modes are arranged so that the pump $\mathrm{TE_0}$ mode attains a group index between those of the signal and idler, thus fulfilling the GVM condition. The use of higher-order modes provides additional dispersion control, enabling robust GVM and improved tolerance to fabrication-induced variations in waveguide width and etch depth (Fig.~S1). Figure~1b illustrates the overall device design. Fabrication details are described in the Methods section of the Supplemental Material. Following the nonlinear interaction region, the circuit directs the signal and idler photons into separate output channels. The idler photon—initially generated in the higher-order $\mathrm{TE_2}$ mode—is subsequently converted to the fundamental $\mathrm{TE_0}$ mode by an integrated mode converter, which exhibits a measured conversion efficiency exceeding 95\%.

To achieve high spectral purities of the generated photon pairs, the JSA should approximate a separable product of two Gaussian functions. This requires that the PEF should have a transform-limited Gaussian shape, while the PMF must likewise assume a Gaussian form. According to Eq.(3), we can modify $g(z)=\chi^{(2)}(z)/\chi_0^{(2)}$ by aperiodically altering the orientation of each domain to obtain a target function\cite{2017Gpdesign,2018independent}. The detailed design flow is provided in Supplementary Material II. In our design, the target cumulative nonlinear amplitude $A_{\mathrm{target}}(z)$ profile along the poled crystal is defined as:
\begin{align}
A_{\mathrm{target}}(z) 
&= C\!\left[
\mathrm{erf}\!\left(\frac{L}{2\sqrt{2}\sigma}\right)
- \mathrm{erf}\!\left(\frac{L - 2z}{2\sqrt{2}\sigma}\right)
\right],
\end{align} which specifies how the effective nonlinearity should grow from the start of the poled region to position $z$. The variable $z$ denotes the position referenced to the poling start ($z=0$), and $L$ is the total poling length. 

The parameter $\sigma$ is the width of the target Gaussian PMF, while $C$ is a normalization constant. By converting $A_{\mathrm{target}}(z)$ into $\phi_{\mathrm{target}}(k)$, we obtain the target PMF shown in Fig.~1c, where different choices of $\sigma$ manifest as varying bandwidths and side-lobe levels. The Gaussian width $\sigma$ governs the trade-off between spectral purity and effective nonlinearity. A narrow $\sigma$ reduces side lobes but lowers brightness, whereas a wider $\sigma$ increases brightness but introduces correlations. We choose $\sigma=L/5$, which minimizes side-lobe formation while maintaining strong nonlinear efficiency. Simulated JSAs further illustrate the impact of this approach. Under the modal interaction conditions shown in Fig. 1a and assuming a matched pump bandwidth, periodic poling produces a sinc-shaped JSA with pronounced spectral correlations (Fig. 1d). These correlations arise from the side lobes of the sinc-like PMF, even though the GVM condition is satisfied. In contrast, the Gaussian-apodized poling implemented on the same waveguide yields a smooth, near-Gaussian, and highly factorable JSA (Fig. 1e), increasing the theoretical spectral purity from 84\% to 99\%. The complete poling configuration used to generate the Gaussian-shaped PMF is shown in Fig.~S2. To experimentally verify that the designed poling profile is accurately implemented on our device, we characterize the poled region using a scanning laser second-harmonic-generation (SHG) microscope images. SHG images (Fig. 1f, right) of the poled region confirm clear, uniform, and well-defined domain inversions (white arrow), with consistent quality observed across multiple regions on the chip (Fig. S3). The dark lines arise from destructive interference between second-harmonic photons generated in neighboring domains, marking the domain boundaries. Our theoretical analysis above shows that combining higher-order-mode dispersion with Gaussian-apodized domain design enables high-purity, co-polarized SSPP generation on thin-film lithium niobate.

\subsection{Device performance measurement}
The PMF of the fabricated device was characterized using sum-frequency generation (SFG), a reverse process of SPDC. Two telecom tunable continuous-wave (cw)  lasers were coupled into separate waveguides (Fig.~2a); one beam was converted to the $\mathrm{TE_2}$ mode through the on-chip mode converter, while the other remained in the fundamental $\mathrm{TE_0}$ mode. When the two beams combined and propagated through the poled region, they satisfied the phase-matching condition and generated SFG light. The SFG signal was collected with a lensed fiber, passed through a wavelength-division multiplexer (WDM), and detected by a silicon photodetector. The measured PMF shows the expected positive slope and a bandwidth in close agreement with simulation (Fig.~2b), confirming both the accuracy of the designed waveguide dispersion and successful realization of the GVM condition. The cutoff in the lower-left corner arises from the bandwidth limit of the idler mode converter. In addition, the SFG image exhibits clear sidelobe suppression, consistent with the Gaussian-apodized poling profile. We quantified the normalized SFG efficiency, defined as $\eta = P_{\mathrm{out}}/(P_1 P_2)$ in the linear regime, obtaining a value of approximately $9\%~\mathrm{W^{-1}}$. Assuming a transform-limited Gaussian pump with optimally tailored bandwidth and center wavelength to match the experimental PMF, the achievable spectral purity of the photon pairs can reach $\sim$95\%.
\\

\begin{figure}
\centering
\includegraphics[width = 0.9\columnwidth]{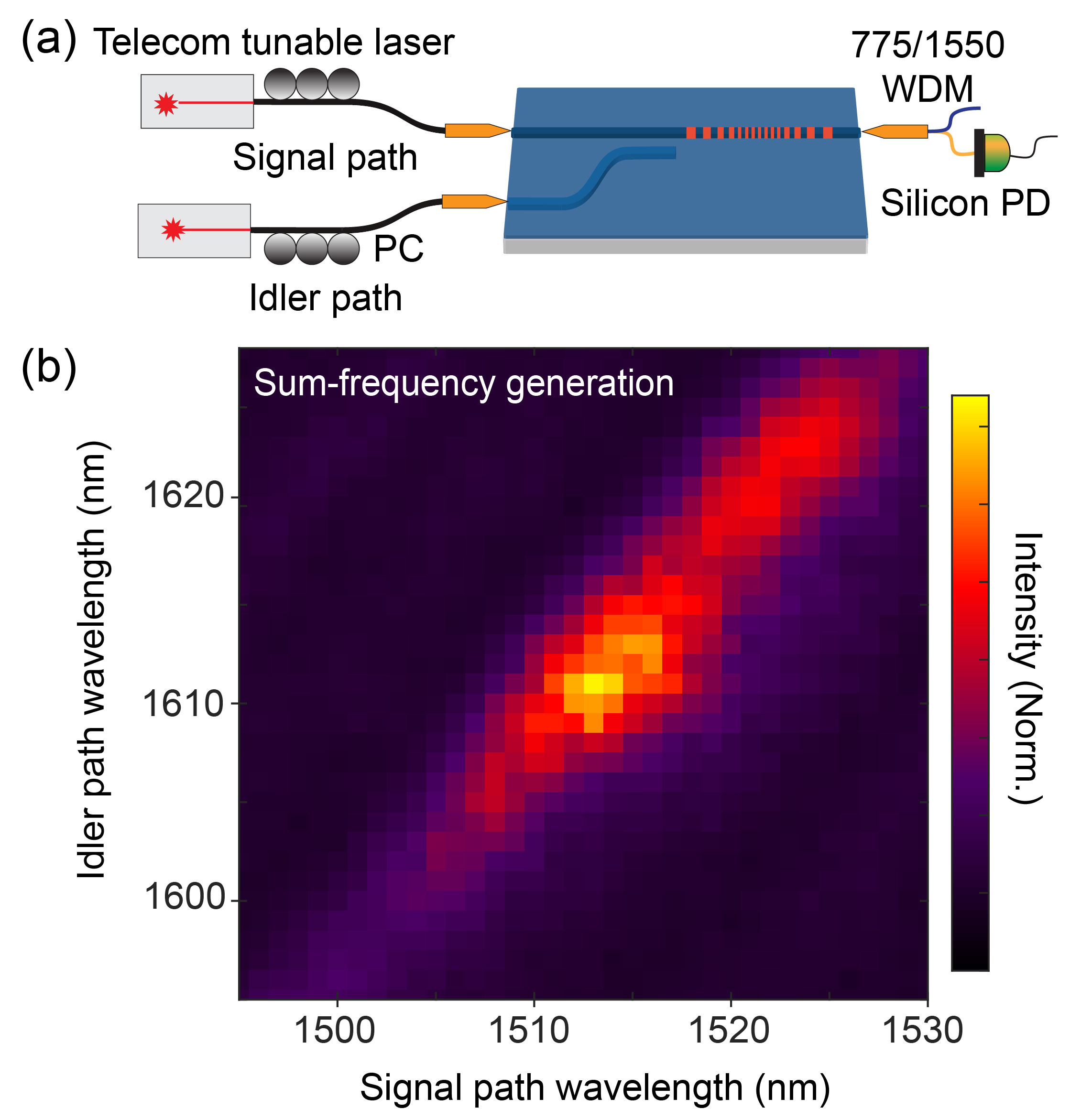}
\caption{\label{Fig. 2}{Experimental characterization of the PMF via SFG process. (a) Schematic of the measurement setup. Two tunable cw telecom lasers are coupled into the device to generate SFG light. One beam is converted into a higher-order mode through on-chip mode converter, while the other remains in the fundamental mode. The generated SFG signal is collected and detected using a silicon photodetector, which is insensitive to the input telecom wavelengths. (b) Measured SFG mapping showing a positively sloped phase-matching response with suppressed sidelobes, in agreement with simulations.}}
\end{figure}

We further evaluated the device performance by measuring its joint spectral intensity (JSI) and unheralded second-order correlation function($g^{(2)}$). Figure 3 shows the experimental setup. To match the pump bandwidth to the device's PMF, the output of an 80-MHz mode-locked femtosecond laser (100-fs pulse duration) was spectrally shaped using a 4-f grating filter (Fig. 3a)\cite{Gaussianmask}. The pump bandwidth was precisely adjusted by a photomask in the Fourier plane that defined the transmission window with a center wavelength of 784 nm and a bandwidth of 4.5 nm. After spectral shaping, the pump beam was split using a beam splitter and detected by a silicon photodiode to serve as the start trigger for the time-correlated single-photon-counting (TCSPC) system. The generated signal and idler photons were guided into separate output waveguides and collected by the two lensed fibers. Residual pump light was blocked using long-pass filters with 1000~nm cutoff wavelength. The outputs were then coupled into two 40~km single-mode fibers, which introduced dispersion for frequency-to-time mapping~\cite{2009Fiber-assisted,chen2017OE}, followed by time-correlated coincidence detection (Fig.~3b). This fiber-dispersion spectrometer allowed us to infer the frequency correlations of the photon pairs from their relative arrival times. The achievable spectral resolution was determined by the fiber length and the timing jitter of the detectors; for standard single-mode fibers (SMF-28, dispersion coefficient $\approx 17~\mathrm{ps/nm/km}$) and superconducting nanowire single-photon detectors (SNSPDs) with 100~ps timing resolution, the effective spectral resolution was about 0.15~nm.

\begin{figure*}[htbp]
\centering
\includegraphics[width = 0.9\textwidth]{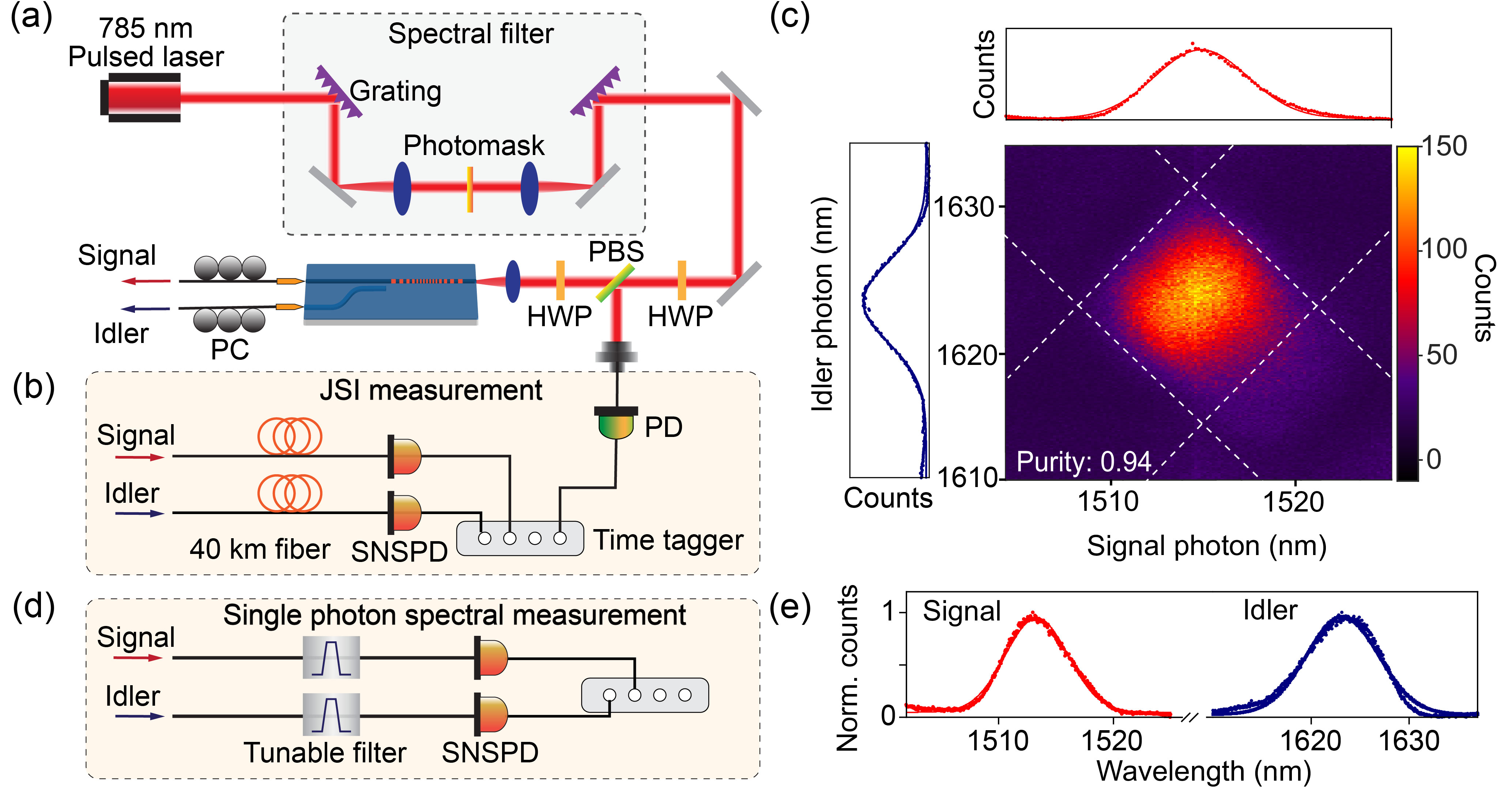}
\caption{\label{Figure3.jpg}Joint-spectral characterization of the co-polarized SSPP. (a) Experimental setup for spectral shaping of the pump light and coupling it into the device. A 4-f grating filter with a photomask at the Fourier plane sets a 4.5-nm bandwidth at a center wavelength of 784 nm. (b) Schematic of the JSI measurement based on a fiber-dispersion spectrometer. The generated signal and idler photons are separated, coupled into two 40-km single-mode fibers for frequency-to-time mapping, and detected by SNSPDs using TCSPC. (c) Reconstructed JSI of the biphoton state, exhibiting a Gaussian profile with strongly suppressed sidelobes, in agreement with both simulations and SFG measurements. (d) Schematic of single-photon spectral measurement using calibrated tunable filters. (e) Independently measured single-photon spectra acquired via calibrated tunable filters, confirming the accuracy of the dispersion-based spectral measurement.}
\end{figure*}

Based on the above setup, we reconstructed the JSI of the biphoton state. The JSI characterizes the spectral correlations between the signal and idler photons and directly reflects their joint spectral distribution\cite{2018JSI}. The measured JSI exhibits a Gaussian profile with strongly suppressed sidelobes (Fig. 3c), in agreement with both numerical simulations and SFG measurements. This consistency confirms the reliability of our source design and demonstrates that the combination of engineered phase-matching and tailored pump bandwidth effectively yields spectrally factorable photon pairs. To verify the JSI shape, we computed the marginal spectral distributions by integrating the JSI over the idler or signal frequency, respectively. Both marginal spectra are well fitted by Gaussian functions, confirming the Gaussian behavior of the biphoton state and the suppression of sidelobes. Building on this result, we performed a Schmidt decomposition, expressed as
\begin{equation}
f(\omega_s,\omega_i) = \sum_{n} \sqrt{\lambda_n}\, u_n(\omega_s)v_n(\omega_i),
\end{equation}
where $\lambda_n$ are the Schmidt coefficients satisfying $\sum_n \lambda_n = 1$, and $u_n(\omega_s)$ and $v_n(\omega_i)$ denote the orthonormal Schmidt modes of the signal and idler photons, respectively. The heralded-state purity is then given by $P = \sum_{n}\lambda_n^2$. From the reconstructed JSI, the Schmidt decomposition yields a Schmidt number of 1.06, corresponding to an estimated purity of $\sim94\%$, confirming that the source operates in a nearly factorable regime. In addition, we independently characterized the single-photon spectra of the photon pairs using a tunable bandpass filter (Fig. 3d). The measured spectral profiles agree with the marginal distributions extracted from the JSI (Fig. 3e), confirming the accuracy of the dispersion-based spectral measurement. 

\begin{figure*}[htbp]
\centering
\includegraphics[width = 0.8\textwidth]{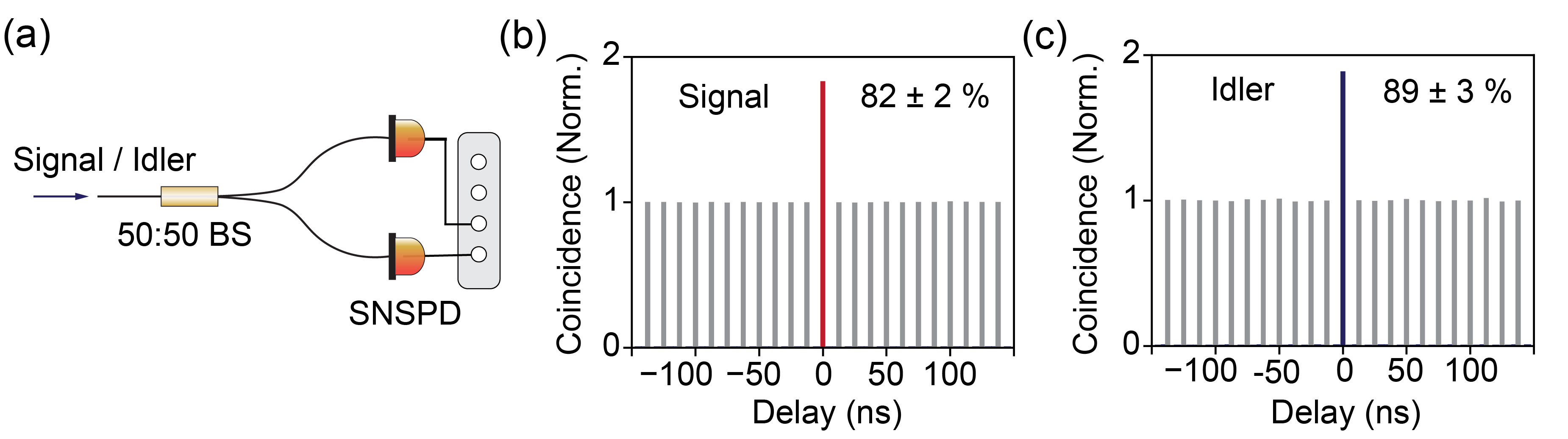}
\caption{\label{fig:4}Second-order correlation measurements of the signal/idler photons. (a) Schematic of measuring the unheralded second-order correlation function $g^{(2)}$ using a Hanbury--Brown--Twiss (HBT) interferometer. (b) Measured coincidence histogram and estimated purity of the signal photons, $P_\text{signal} = 82 \pm 2\%$. (c) Measured coincidence histogram and estimated purity of the idler photons, $P_\text{idler} = 89 \pm 3\%$.}
\end{figure*}

To further verify and quantify the separability of co-polarized SSPP, we measured the unheralded second-order correlation $g^{(2)}$. While the JSI captures only intensity-level correlations, the $g^{(2)}$ is sensitive to phase-dependent correlations in the joint spectral amplitude\cite{christ2011probing}, as well as residual entanglement in other degrees of freedom. The signal and idler photons were independently isolated using 12-nm band-pass filters, with bandwidths exceeding the intrinsic photon spectral widths. The signal arm was then split by a 50:50 fiber coupler and detected by two SNSPDs to form a Hanbury--Brown--Twiss interferometer (Fig.~4a). A 5~ns coincidence window was used to ensure that the normalized zero-delay coincidence counts directly yielded $g^{(2)}(\Delta T)$, from which the heralded-state purity can be estimated using $P \approx \frac{g^{(2)}(0)}{g^{(2)}(\infty)} - 1$. Similar measurement was performed for the idler arm. Based on the unheralded $g^{(2)}$ measurement, the estimated purities for the signal and idler photons were $82\%\pm2\%$ and $89\%\pm3\%$, respectively (Figs.~4b--c). While GVM ensures a separable JSI, residual phase correlations---though not visible in the intensity distribution---may persist in the JSA due to the combined effect of the Gaussian pump envelope and the PMF, thereby reducing heralded purity and interference visibility\cite{JSAphase}. Imperfections in the poling structure can further introduce unwanted spectral correlations, lowering the overall factorability of the generated photon pairs.

\section{Conclusions}
In summary, the modal-dispersion approach establishes a flexible and broadly applicable framework for phase-matching engineering, breaking the long-standing limitations of conventional type-II configurations for generation of spectrally separable photon pairs. Whereas phase matching in type-II schemes relies on the dispersion of cross-polarized modes and is therefore intrinsically constrained, higher-order modes enable precise tuning of group-velocity relations and dispersion curvature, providing high design freedom for realizing factorable joint spectral amplitudes and engineered spectral envelopes. Importantly, this strategy operates entirely within a co-polarized configuration, which substantially simplifies photonic circuit design and eliminates the need for polarization management. This enhanced design freedom is particularly valuable not only for tailoring the spectral properties of integrated quantum sources, but also for enabling scalable architectures in which photon generation, routing, and spectral manipulation can be co-designed. More broadly, it facilitates seamless integration with a wide range of photonic and hybrid quantum systems, paving the way toward more versatile and functionally rich platforms for large-scale quantum technologies.

\begin{acknowledgments}
We thank Changchen Chen for helpful discussion. This research is supported by the National Research Foundation Singapore (NRF-NRFF15-2023-0005), Singapore Ministry of Education (MOET32024-0009), and Centre for Quantum Technologies Funding Initiative (S24Q2d0009).    
\end{acknowledgments}
\section*{Author Declarations}
The authors have no conflicts of interest.

\section*{Author Contribution} 
D.Z., X.W. conceived the idea. X.W designed the devices. X.W, S.S.W, Z.Y., X.C., G.W., H.H, and V.D. fabricated the devices. X.W., L.Z., Y.L., X.S., R.Y., and S.W. performed the measurements of the devices. D.Z. supervised the project. All the authors discussed the results and wrote the manuscript.

\section*{Data Availability Statement}
The data that support the findings of this study are available from the corresponding author upon reasonable request.


\bibliography{apssamp}

@article{psiquantum2025,
    author={PsiQuanutm team},
  title={A manufacturable platform for photonic quantum computing},
  journal={Nature},
  volume={641},
  number={8064},
  pages={876--883},
  year={2025},
  publisher={Nature Publishing Group UK London}
}

@article{zhong2020quantum,
  title={Quantum computational advantage using photons},
  author={Zhong, Han-Sen and Wang, Hui and Deng, Yu-Hao and Chen, Ming-Cheng and Peng, Li-Chao and Luo, Yi-Han and Qin, Jian and Wu, Dian and Ding, Xing and Hu, Yi and others},
  journal={Science},
  volume={370},
  number={6523},
  pages={1460--1463},
  year={2020},
  publisher={American Association for the Advancement of Science}
}

@article{kok2007linear,
  title={Linear optical quantum computing with photonic qubits},
  author={Kok, Pieter and Munro, William J and Nemoto, Kae and Ralph, Timothy C and Dowling, Jonathan P and Milburn, Gerard J},
  journal={Reviews of modern physics},
  volume={79},
  number={1},
  pages={135--174},
  year={2007},
  publisher={APS}
}

@article{zhudilight,
  title={Spectral control of nonclassical light pulses using an integrated thin-film lithium niobate modulator},
  author={Zhu, Di and Chen, Changchen and Yu, Mengjie and Shao, Linbo and Hu, Yaowen and Xin, CJ and Yeh, Matthew and Ghosh, Soumya and He, Lingyan and Reimer, Christian and others},
  journal={Light: Science \& Applications},
  volume={11},
  number={1},
  pages={327},
  year={2022},
  publisher={Nature Publishing Group UK London}
}

@article{on-chip-sources,
  title={On-chip heralded single photon sources},
  author={Signorini, S and Pavesi, L},
  journal={AVS Quantum Science},
  volume={2},
  number={4},
  pages={041701},
  year={2020},
  publisher={AIP Publishing}
}

@article{zhudireview,
  title={Integrated photonics on thin-film lithium niobate},
  author={Zhu, Di and Shao, Linbo and Yu, Mengjie and Cheng, Rebecca and Desiatov, Boris and Xin, C\_J and Hu, Yaowen and Holzgrafe, Jeffrey and Ghosh, Soumya and Shams-Ansari, Amirhassan and others},
  journal={Advances in Optics and Photonics},
  volume={13},
  number={2},
  pages={242--352},
  year={2021},
  publisher={Optical Society of America}
}

@article{2019photonic,
  title={Photonic quantum information processing: A concise review},
  author={Slussarenko, Sergei and Pryde, Geoff J},
  journal={Applied physics reviews},
  volume={6},
  number={4},
  pages={041303},
  year={2019},
  publisher={AIP Publishing}
}

@article{xin2022,
  title={Spectrally separable photon-pair generation in dispersion engineered thin-film lithium niobate},
  author={Xin, C\_J and Mishra, Jatadhari and Chen, Changchen and Zhu, Di and Shams-Ansari, Amirhassan and Langrock, Carsten and Sinclair, Neil and Wong, Franco NC and Fejer, M\_M and Lon{\v{c}}ar, Marko},
  journal={Optics Letters},
  volume={47},
  number={11},
  pages={2830--2833},
  year={2022},
  publisher={Optica Publishing Group}
}

@article{madsen2022quantum,
  title={Quantum computational advantage with a programmable photonic processor},
  author={Madsen, Lars S and Laudenbach, Fabian and Askarani, Mohsen Falamarzi and Rortais, Fabien and Vincent, Trevor and Bulmer, Jacob FF and Miatto, Filippo M and Neuhaus, Leonhard and Helt, Lukas G and Collins, Matthew J and others},
  journal={Nature},
  volume={606},
  number={7912},
  pages={75--81},
  year={2022},
  publisher={Nature Publishing Group UK London}
}

@article{flamini2018photonic,
  title={Photonic quantum information processing: a review},
  author={Flamini, Fulvio and Spagnolo, Nicolo and Sciarrino, Fabio},
  journal={Reports on Progress in Physics},
  volume={82},
  number={1},
  pages={016001},
  year={2018},
  publisher={IOP Publishing}
}

@article{2025counter,
  title={Counter-propagating spontaneous parametric down-conversion source in lithium niobate on insulator},
  author={Kellner, Jost and Sabatti, Alessandra and Kuttner, Tristan and Chapman, Robert J and Grange, Rachel},
  journal={arXiv preprint arXiv:2506.21396},
  year={2025}
}

@article{2025typeii,
  title={Scalable quantum interference in integrated lithium niobate nanophotonics},
  author={Kuttner, Tristan and Sabatti, Alessandra and Kellner, Jost and Grange, Rachel and Chapman, Robert J},
  journal={arXiv preprint arXiv:2506.20519},
  year={2025}
}

@book{multiphoton,
  title={Multi-photon quantum interference},
  author={Ou, Zhe-Yu Jeff},
  volume={43},
  year={2007},
  publisher={Springer}
}

@article{2021counter,
  title={Observation of frequency-uncorrelated photon pairs generated by counter-propagating spontaneous parametric down-conversion},
  author={Liu, Yi-Chen and Guo, Dong-Jie and Ren, Kun-Qian and Yang, Ran and Shang, Minghao and Zhou, Wei and Li, Xinhui and Sun, Chang-Wei and Xu, Ping and Xie, Zhenda and others},
  journal={Scientific Reports},
  volume={11},
  number={1},
  pages={12628},
  year={2021},
  publisher={Nature Publishing Group UK London}
}

@article{2017PRALimits,
  title={Limits on the heralding efficiencies and spectral purities of spectrally filtered single photons from photon-pair sources},
  author={Meyer-Scott, Evan and Montaut, Nicola and Tiedau, Johannes and Sansoni, Linda and Herrmann, Harald and Bartley, Tim J and Silberhorn, Christine},
  journal={Physical Review A},
  volume={95},
  number={6},
  pages={061803},
  year={2017},
  publisher={APS}
}

@article{2017PRLheralded,
  title={Heralded single photons based on spectral multiplexing and feed-forward control},
  author={Grimau Puigibert, M and Aguilar, GH and Zhou, Q and Marsili, F and Shaw, MD and Verma, VB and Nam, SW and Oblak, D and Tittel, W},
  journal={Physical Review Letters},
  volume={119},
  number={8},
  pages={083601},
  year={2017},
  publisher={APS}
}

@article{2008PRLwal,
  title={Heralded generation of ultrafast single photons in pure quantum states},
  author={Mosley, Peter J and Lundeen, Jeff S and Smith, Brian J and Wasylczyk, Piotr and U’Ren, Alfred B and Silberhorn, <? format?> Christine and Walmsley, Ian A},
  journal={Physical Review Letters},
  volume={100},
  number={13},
  pages={133601},
  year={2008},
  publisher={APS}
}

@article{HOM1987,
  title={Measurement of subpicosecond time intervals between two photons by interference},
  author={Hong, Chong-Ki and Ou, Zhe-Yu and Mandel, Leonard},
  journal={Physical review letters},
  volume={59},
  number={18},
  pages={2044},
  year={1987},
  publisher={APS}
}

@article{2001PRAeliminating,
  title={Eliminating frequency and space-time correlations in multiphoton states},
  author={Grice, Warren P and U’Ren, Alfred B and Walmsley, Ian A},
  journal={Physical Review A},
  volume={64},
  number={6},
  pages={063815},
  year={2001},
  publisher={APS}
}

@article{2018design,
  title={Design considerations for high-purity heralded single-photon sources},
  author={Graffitti, Francesco and Kelly-Massicotte, J{\'e}r{\'e}my and Fedrizzi, Alessandro and Bra{\'n}czyk, Agata M},
  journal={Physical Review A},
  volume={98},
  number={5},
  pages={053811},
  year={2018},
  publisher={APS}
}

@article{2019SA,
  title={Nonlinear integrated quantum electro-optic circuits},
  author={Luo, Kai-Hong and Brauner, Sebastian and Eigner, Christof and Sharapova, Polina R and Ricken, Raimund and Meier, Torsten and Herrmann, Harald and Silberhorn, Christine},
  journal={Science advances},
  volume={5},
  number={1},
  pages={eaat1451},
  year={2019},
  publisher={American Association for the Advancement of Science}
}

@article{chen2017OE,
  title={Efficient generation and characterization of spectrally factorable biphotons},
  author={Chen, Changchen and Bo, Cao and Niu, Murphy Yuezhen and Xu, Feihu and Zhang, Zheshen and Shapiro, Jeffrey H and Wong, Franco NC},
  journal={Optics express},
  volume={25},
  number={7},
  pages={7300--7312},
  year={2017},
  publisher={Optical Society of America}
}

@article{APR2025nanodomain,
  title={Nonlinear domain engineering for quantum technologies},
  author={Weiss, Tim F and Peruzzo, Alberto},
  journal={Applied Physics Reviews},
  volume={12},
  number={1},
  pages={011318},
  year={2025},
  publisher={AIP Publishing}
}

@article{2017Gpdesign,
  title={Pure down-conversion photons through sub-coherence-length domain engineering},
  author={Graffitti, Francesco and Kundys, Dmytro and Reid, Derryck T and Bra{\'n}czyk, Agata M and Fedrizzi, Alessandro},
  journal={Quantum Science and Technology},
  volume={2},
  number={3},
  pages={035001},
  year={2017},
  publisher={IOP Publishing}
}

@article{2018independent,
  title={Independent high-purity photons created in domain-engineered crystals},
  author={Graffitti, Francesco and Barrow, Peter and Proietti, Massimiliano and Kundys, Dmytro and Fedrizzi, Alessandro},
  journal={Optica},
  volume={5},
  number={5},
  pages={514--517},
  year={2018},
  publisher={Optical Society of America}
}

@article{2020PRL,
  title={High quality entangled photon pair generation in periodically poled thin-film lithium niobate waveguides},
  author={Zhao, Jie and Ma, Chaoxuan and R{\"u}sing, Michael and Mookherjea, Shayan},
  journal={Physical review letters},
  volume={124},
  number={16},
  pages={163603},
  year={2020},
  publisher={APS}
}

@article{wangcoptica2018,
  title={Ultrahigh-efficiency wavelength conversion in nanophotonic periodically poled lithium niobate waveguides},
  author={Wang, Cheng and Langrock, Carsten and Marandi, Alireza and Jankowski, Marc and Zhang, Mian and Desiatov, Boris and Fejer, Martin M and Lon{\v{c}}ar, Marko},
  journal={Optica},
  volume={5},
  number={11},
  pages={1438--1441},
  year={2018},
  publisher={Optical Society of America}
}

@article{1997spectral,
  title={Spectral information and distinguishability in type-II down-conversion with a broadband pump},
  author={Grice, Warren P and Walmsley, Ian A},
  journal={Physical Review A},
  volume={56},
  number={2},
  pages={1627},
  year={1997},
  publisher={APS}
}

@article{mosley2008conditional,
  title={Conditional preparation of single photons using parametric downconversion: a recipe for purity},
  author={Mosley, Peter J and Lundeen, Jeff S and Smith, Brian J and Walmsley, Ian A},
  journal={New Journal of Physics},
  volume={10},
  number={9},
  pages={093011},
  year={2008},
  publisher={IOP Publishing}
}

@article{2018heralding,
  title={Heralding pure single photons: A comparison between counterpropagating and copropagating twin photons},
  author={Gatti, Alessandra and Brambilla, Enrico},
  journal={Physical Review A},
  volume={97},
  number={1},
  pages={013838},
  year={2018},
  publisher={APS}
}

@article{2023pranarrow,
  title={Narrow-band photon pair generation through cavity-enhanced spontaneous parametric down-conversion},
  author={Mataji-Kojouri, Amideddin and Liscidini, Marco},
  journal={Physical Review A},
  volume={108},
  number={5},
  pages={053714},
  year={2023},
  publisher={APS}
}

@article{PRL2020ultrabright,
  title={Ultrabright quantum photon sources on chip},
  author={Ma, Zhaohui and Chen, Jia-Yang and Li, Zhan and Tang, Chao and Sua, Yong Meng and Fan, Heng and Huang, Yu-Ping},
  journal={Physical Review Letters},
  volume={125},
  number={26},
  pages={263602},
  year={2020},
  publisher={APS}
}

@article{2006generation,
  title={Generation of pure-state single-photon wavepackets by conditional preparation based on spontaneous parametric downconversion},
  author={U'Ren, Alfred B and Silberhorn, Christine and Erdmann, Reinhard and Banaszek, Konrad and Grice, Warren P and Walmsley, Ian A and Raymer, Michael G},
  journal={arXiv preprint quant-ph/0611019},
  year={2006}
}

@article{2009Fiber-assisted,
  title={Fiber-assisted single-photon spectrograph},
  author={Avenhaus, Malte and Eckstein, Andreas and Mosley, Peter J and Silberhorn, Christine},
  journal={Optics letters},
  volume={34},
  number={18},
  pages={2873--2875},
  year={2009},
  publisher={Optical Society of America}
}

@article{Gaussianmask,
  title={Generation of optical coherent-state superpositions by number-resolved photon subtraction from the squeezed vacuum},
  author={Gerrits, Thomas and Glancy, Scott and Clement, Tracy S and Calkins, Brice and Lita, Adriana E and Miller, Aaron J and Migdall, Alan L and Nam, Sae Woo and Mirin, Richard P and Knill, Emanuel},
  journal={Physical Review A—Atomic, Molecular, and Optical Physics},
  volume={82},
  number={3},
  pages={031802},
  year={2010},
  publisher={APS}
}

@article{christ2011probing,
  title={Probing multimode squeezing with correlation functions},
  author={Christ, Andreas and Laiho, Kaisa and Eckstein, Andreas and Cassemiro, Kati{\'u}scia N and Silberhorn, Christine},
  journal={New Journal of Physics},
  volume={13},
  number={3},
  pages={033027},
  year={2011},
  publisher={IOP Publishing}
}

@article{JSAphase,
  title={Phase-sensitive tomography of the joint spectral amplitude of photon pair sources},
  author={Jizan, Iman and Bell, Bryn and Helt, Lukas G and Bedoya, Alvaro Casas and Xiong, Chunle and Eggleton, Benjamin J},
  journal={Optics letters},
  volume={41},
  number={20},
  pages={4803--4806},
  year={2016},
  publisher={Optical Society of America}
}

@article{2018JSI,
  title={Joint spectral characterization of photon-pair sources},
  author={Zielnicki, Kevin and Garay-Palmett, Karina and Cruz-Delgado, Daniel and Cruz-Ramirez, Hector and O’Boyle, Michael F and Fang, Bin and Lorenz, Virginia O and U’Ren, Alfred B and Kwiat, Paul G},
  journal={Journal of Modern Optics},
  volume={65},
  number={10},
  pages={1141--1160},
  year={2018},
  publisher={Taylor \& Francis}
}

\end{document}



\title{On-Chip Generation of Co-Polarized and Spectrally Separable Photon Pairs 
}

\author{Xiaojie Wang}
\affiliation{Department of Materials Science and Engineering, National University of Singapore, 117575, Singapore}
\affiliation{Centre for Quantum Technologies, National University of Singapore, 117543, Singapore}

\author{Lin Zhou}%
\affiliation{Department of Materials Science and Engineering, National University of Singapore, 117575, Singapore}
\affiliation{Centre for Quantum Technologies, National University of Singapore, 117543, Singapore}

\author{Yue Li}%
\affiliation{Department of Materials Science and Engineering, National University of Singapore, 117575, Singapore}

\author{Sakthi Sanjeev Mohanraj}%
\affiliation{Department of Materials Science and Engineering, National University of Singapore, 117575, Singapore}

\author{Xiaodong Shi}
\affiliation{A$^\ast$STAR Quantum Innovation Centre (Q.InC), Agency for Science, Technology and Research (A$^\ast$STAR), 138634, Singapore}%
\affiliation{Institute of Materials Research and Engineering (IMRE), Agency for Science, Technology and Research (A$^\ast$STAR), 138634, Singapore}%

\author{Zhuoyang Yu}%
\affiliation{Department of Materials Science and Engineering, National University of Singapore, 117575, Singapore}
\affiliation{Centre for Quantum Technologies, National University of Singapore, 117543, Singapore}


\author{\\Ran Yang}%
\affiliation{Department of Materials Science and Engineering, National University of Singapore, 117575, Singapore}

\author{Xu Chen}%
\affiliation{Department of Materials Science and Engineering, National University of Singapore, 117575, Singapore}

\author{Guangxing Wu}%
\affiliation{Department of Materials Science and Engineering, National University of Singapore, 117575, Singapore}
\affiliation{Centre for Quantum Technologies, National University of Singapore, 117543, Singapore}

\author{Hao Hao}
\affiliation{Centre for Quantum Technologies, National University of Singapore, 117543, Singapore}

\author{Sihao Wang}
\affiliation{A$^\ast$STAR Quantum Innovation Centre (Q.InC), Agency for Science, Technology and Research (A$^\ast$STAR), 138634, Singapore}%
\affiliation{Institute of Materials Research and Engineering (IMRE), Agency for Science, Technology and Research (A$^\ast$STAR), 138634, Singapore}%

\author{Veerendra Dhyani}
\affiliation{A$^\ast$STAR Quantum Innovation Centre (Q.InC), Agency for Science, Technology and Research (A$^\ast$STAR), 138634, Singapore}%
\affiliation{Institute of Materials Research and Engineering (IMRE), Agency for Science, Technology and Research (A$^\ast$STAR), 138634, Singapore}%

\author{Di Zhu}
\email{dizhu@nus.edu.sg}
\affiliation{Department of Materials Science and Engineering, National University of Singapore, 117575, Singapore}
\affiliation{Centre for Quantum Technologies, National University of Singapore, 117543, Singapore}
\affiliation{A$^\ast$STAR Quantum Innovation Centre (Q.InC), Agency for Science, Technology and Research (A$^\ast$STAR), 138634, Singapore}
\affiliation{Institute of Materials Research and Engineering (IMRE), Agency for Science, Technology and Research (A$^\ast$STAR), 138634, Singapore}

\title{Supplementary Materials for\\
        On-Chip Generation of Co-Polarized and Spectrally Separable Photon Pairs}
\pacs{} \maketitle
\onecolumngrid





\section{Material and Methods}
\subsection{Device design}
The nanophotonic waveguide is designed along the crystal $y$-axis on a 600-nm-thick $x$-cut thin-film lithium niobate (TFLN) platform on insulator substrate. 
The device fully exploits the large second-order nonlinear coefficient $d_{33}$ of lithium niobate. The ridge waveguide has a width of 2.0 $\mu$m and an etching depth of 360~nm, optimized to support the fundamental ($\mathrm{TE}_0$) for signal photon and higher-order ($\mathrm{TE}_2$) modes for idler photon. 
The phase-matching period for the type-0 spontaneous parametric down-conversion (SPDC) process is determined by $\Lambda = \frac{2\pi}{k_p - k_s - k_i}$ where $k_{p,s,i}$ are the wavevectors and effective refractive indices of the pump, signal, and idler modes, respectively. Based on the simulated effective indices near 785 nm (pump), 1520 nm (signal) and 1620 nm (idler), we obtain an optimal period of $\Lambda = 3.08~\mu$m and a total poling length of about 9.2 mm. 

\subsection{Device fabrication}
The waveguides were fabricated on an x-cut MgO-doped thin-film lithium niobate (TFLN) substrate consisting of a 600~nm lithium niobate layer. The periodic poling was carried out prior to waveguide patterning. A layer of polymethyl methacrylate (PMMA) resist was first spin-coated onto the TFLN chip and patterned using electron-beam lithography (EBL) to define the electrode geometry. Subsequently, a 100~nm thick nickel (Ni) layer was deposited by electron-beam evaporation, followed by lift-off to form comb-shaped electrodes for periodic poling. Electrical poling was performed by applying a sequence of voltage pulses opposite to the intrinsic ferroelectric polarization of the crystal, leading to controlled domain inversion within the designed region. After poling, the waveguide patterns were defined using a negative-tone ma-N 2405 resist by EBL, and the features were transferred into the TFLN layer with an etching depth of 360 nm through inductively coupled plasma reactive-ion etching (ICP-RIE) with argon gas. Finally, a SiO$_2$ cladding layer was deposited by plasma-enhanced chemical vapor deposition (PECVD) to protect the surface and ensure long-term device stability.

\section{Simulation of Group Index}
\par We simulate the group indices of the pump ($\mathrm{TE_0}$), signal ($\mathrm{TE_0}$), and idler ($\mathrm{TE_2}$) modes for waveguides with different geometric parameters. These simulated results show how waveguide geometry shapes modal dispersion and determine whether the group-velocity-matching (GVM) condition can be satisfied for all three interacting fields. Achieving GVM requires that the pump's group index lie between those of the signal and idler. The results reveal that the required group-velocity ordering can be compactly expressed as $\left(v_{g,s}^{-1}-v_{g,p}^{-1}\right)/\left(v_{g,p}^{-1}-v_{g,i}^{-1}\right) \ge 0$,
which ensures that the pump’s inverse group velocity lies between those of the signal and idler. This condition includes boundary cases where two group velocities become equal. The condition can be written in terms of group indices as either $n_{g,s} \le n_{g,p} < n_{g,i}$ or $n_{g,s} < n_{g,p} \le n_{g,i}$, 
\begin{figure}[ht]
    \centering
    \includegraphics[width=0.8\textwidth]{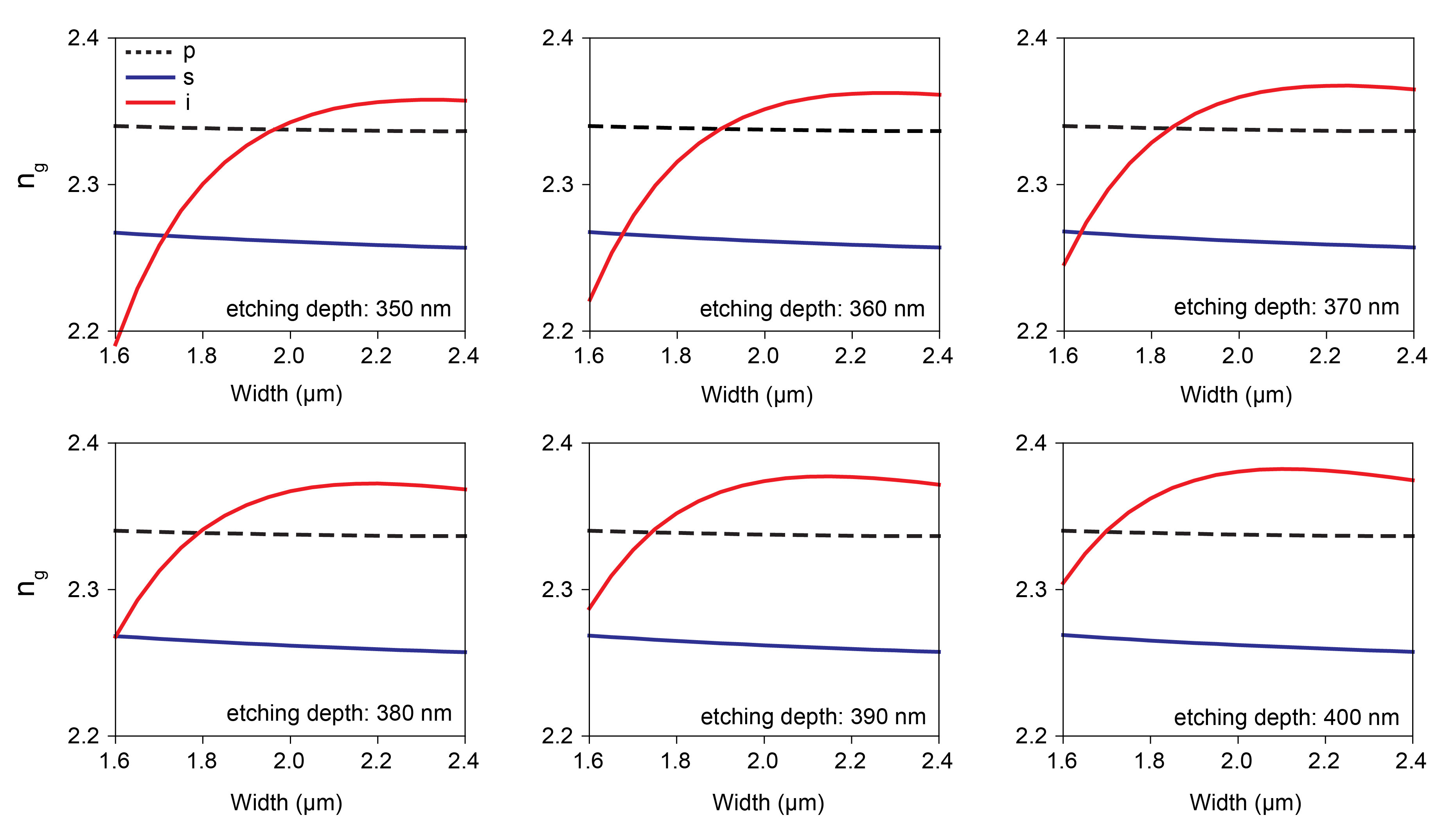}
    \caption{Simulated group index of the pump, signal, and idler photons under different waveguide parameters.}
    \label{fig:S1}
    \end{figure}

\section{Design for Gaussian PMF}
\par To realize a nearly factorable biphoton state, we design the crystal’s nonlinear
domain structure to generate a Gaussian-shaped phase-matching function (PMF).
We define the phase mismatch as
\begin{equation}
\Delta k = k_p - k_s - k_i - \frac{2\pi}{\Lambda},
\end{equation}
where $k_p$, $k_s$, and $k_i$ are the pump, signal, and idler wavevectors,
and $\Lambda$ is the quasi-phase-matching period.
The target PMF in the spatial-frequency domain is chosen as
\begin{equation}
\phi_{\mathrm{target}}(\Delta k)
= \exp\!\left(-\frac{\sigma^{2}}{2}\Delta k^{2}\right),
\end{equation}
where $\sigma$ controls the bandwidth of the target PMF.

Performing an inverse Fourier transform yields the corresponding target nonlinear
profile in the spatial domain. For a crystal of finite length $L$, the Gaussian
profile is centered at $z=L/2$:
\begin{equation}
\Phi_{\mathrm{target}}(z)
= \frac{1}{\sqrt{2\pi}\sigma}
\exp\!\left[-\frac{(z - L/2)^{2}}{2\sigma^{2}}\right],
\end{equation}
where $z$ denotes the longitudinal position along the poled crystal.

The cumulative nonlinear amplitude generated from the poling start position
$(z=0)$ to a position $z$ is obtained by integrating
$\Phi_{\mathrm{target}}(z)$:
\begin{align}
A_{\mathrm{target}}(z)
&= C\int_{0}^{z}
\frac{1}{\sqrt{2\pi}\sigma}
\exp\!\left[-\frac{(z' - L/2)^{2}}{2\sigma^{2}}\right]
dz' \notag \\
&= C\left[
\mathrm{erf}\!\left(\frac{L}{2\sqrt{2}\sigma}\right)
-
\mathrm{erf}\!\left(\frac{L - 2z}{2\sqrt{2}\sigma}\right)
\right],
\end{align}
where $C$ is a normalization constant.

For a periodically poled structure with binary domain orientation
$g(z)=\pm1$, the PMF is given by
\begin{equation}
\phi(\Delta k)
= \int_{0}^{L} g(z)\, e^{i\Delta k z}\, dz .
\end{equation}
The contribution of the $j$th domain of length $L_c$ is
\begin{equation}
\int_{(j-1)L_c}^{jL_c} g[j]\, e^{i\Delta k z'}\, dz'
= g[j]\,
\frac{e^{i\Delta k j L_c}-e^{i\Delta k (j-1)L_c}}{i\Delta k}.
\end{equation}
Summing over all domains yields the discrete PMF
\begin{equation}
\phi(\Delta k)
= \sum_{j=1}^{N}
g[j]\,
\frac{e^{i\Delta k j L_c}-e^{i\Delta k (j-1)L_c}}{i\Delta k},
\qquad
N = \frac{L}{L_c}.
\end{equation}
Equivalently,
\begin{equation}
\phi(\Delta k)
= \frac{e^{i\Delta k L_c}-1}{i\Delta k}
\sum_{j=1}^{N} g[j]\, e^{i\Delta k (j-1)L_c},
\end{equation}
where $(e^{i\Delta k L_c}-1)/(i\Delta k)$ represents the contribution of a
single domain, and the summation describes the coherent accumulation of all
domains.

To determine the binary domain sequence $g[j]=\pm1$, we employ a
cumulative-error minimization algorithm. Each domain orientation is chosen
such that the discrete cumulative nonlinear amplitude
\begin{equation}
A[j] = \sum_{m=1}^{j} g[m]\,L_c
\end{equation}
best approximates the target function
$A_{\mathrm{target}}(jL_c)$.
This procedure produces a domain sequence whose resulting PMF closely
approximates the desired Gaussian profile.
The biphoton joint spectral amplitude (JSA) can be expressed via Schmidt
decomposition as
\begin{equation}
f(\omega_s,\omega_i)
= \sum_{n} c_n\,
\chi_n(\omega_s)\,
\varphi_n(\omega_i),
\end{equation}
where $\chi_n$ and $\varphi_n$ are orthonormal mode functions and
$\sum_n |c_n|^2 = 1$.
The spectral purity is given by
\begin{equation}
P = \sum_n |c_n|^4,
\end{equation}
with $P=1$ corresponding to a fully factorable biphoton state.

\begin{figure}[ht]
\centering
\includegraphics[width =0.8\columnwidth]{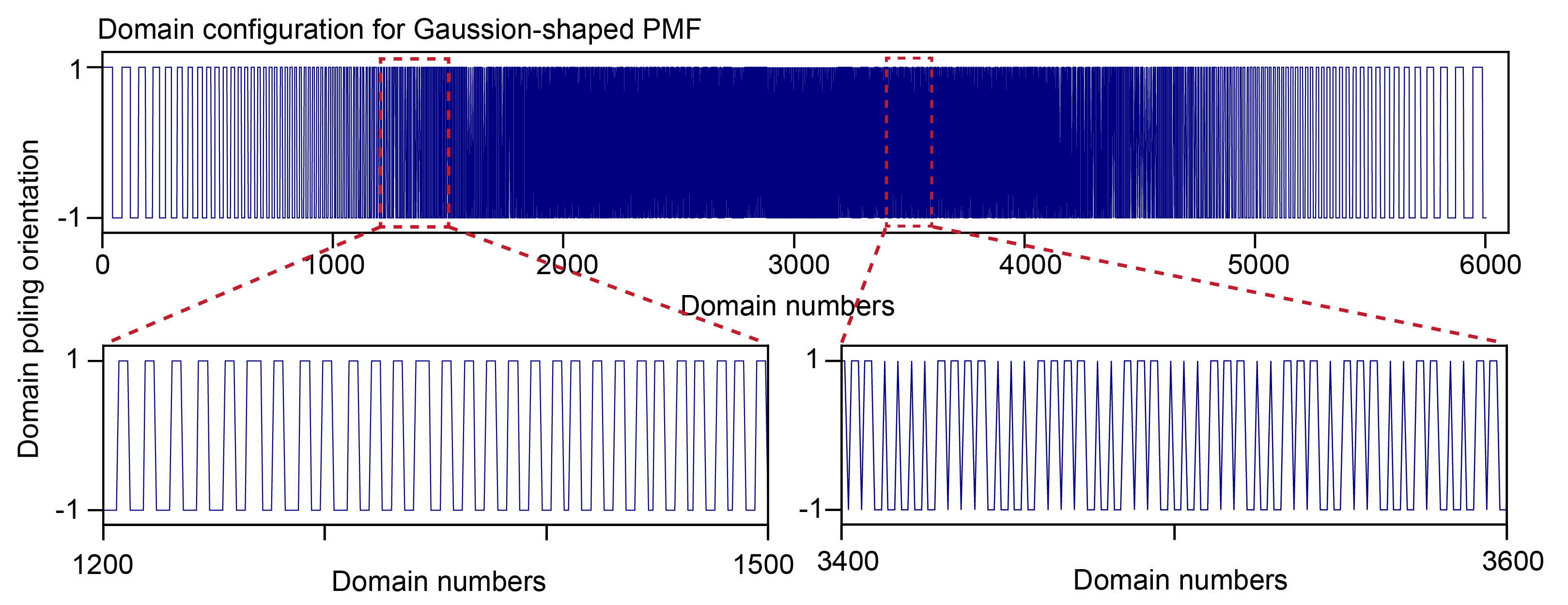}
\caption{The poling domain distribution $g(z)$, representing the spatial modulation of the nonlinear coefficient}
\label{fig:S2}
\end{figure}


\section{Second-harmonic generation (SHG) imaging of device}
\par The quality of the poled regions on the device was examined using SHG microscopy.
As shown in Fig. S3, the bright areas correspond to regions with strong SHG response, while the dark lines mark the reconstructed domain walls, clearly outlining the boundaries between the initially polarized domains and the electrically reversed domains.The uniform stripe spacing and consistent contrast across multiple segments confirm clean and well-defined domain reversal, demonstrating the high repeatability of the poling process.
\vspace{2pt}
\begin{figure}[htbp]
\centering
\includegraphics[width =0.45\columnwidth]{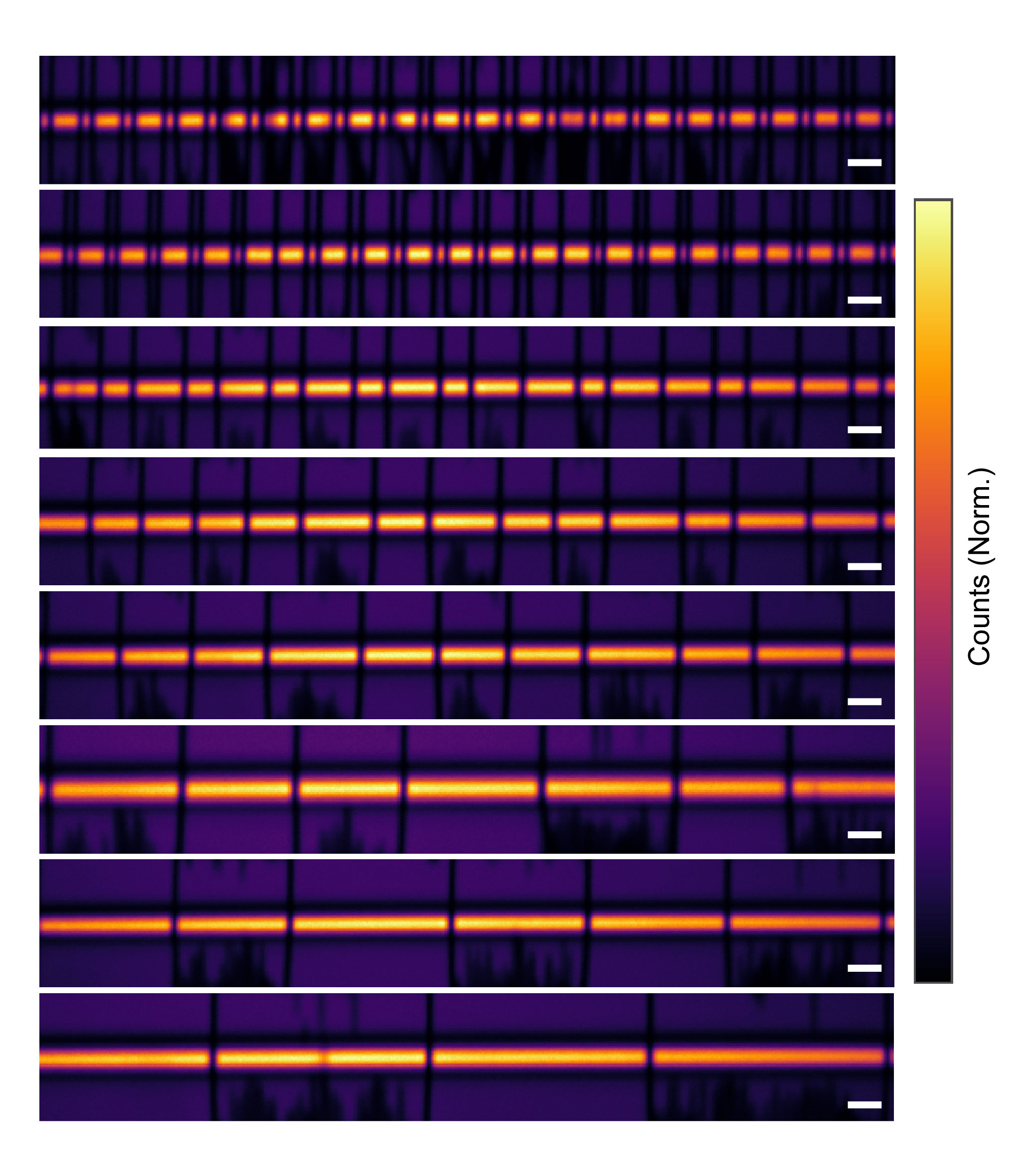}
\caption{SHG images of multiple poled regions. Scale bar: 5~\textmu m.}
\label{fig:S3}
\end{figure}